\documentclass[aps,twocolumn,showpacs,preprintnumbers,amsmath,amssymb,floatfix]{revtex4}

\usepackage{graphicx}
\usepackage{dcolumn}
\usepackage{bm}
\usepackage[german,american]{babel}
\newcommand{\var}[1]{\ensuremath{#1}}
\newcommand{\sep}{\ensuremath{\sigma_{\epsilon}}}
\newcommand{\Gcat}{\ensuremath{\Gamma_{cat}}}
\newcommand{\Gcatmax}{\ensuremath{\Gamma_{cat,max}}}
\newcommand{\Gan}{\ensuremath{\Gamma_{stat}}}
\newcommand{\s}{\ensuremath{\sigma_{r}}}
\newcommand{\kbt}{\ensuremath{k_{B}T}}
\newcommand{\vc}{\ensuremath{V_{c}}}
\newcommand{\rs}{\ensuremath{r_{s}}}

\newcommand{\rem}{REM}
\newcommand{\cim}{CIM}
\newcommand{\ddc}{\ensuremath{D_{DC}}}
\newcommand{\dac}{\ensuremath{D_{AC}}}
\newcommand{\figref}[1]{Fig.\ \ref{#1}}
\newcommand{\ded}{\ensuremath{\Delta E_{dynamic}}}
\newcommand{\des}{\ensuremath{\Delta E_{static}}}
\newcommand{\eqnref}[1]{eqn.\ (\ref{#1})}
\newcommand{\ea}{\ensuremath{E_{a}}}
\newcommand{\rsdt}{\mbox{\ensuremath{\langle r^{2}(t)\rangle/t}}}
\newcommand{\gofr}{\ensuremath{g(r)}}

\begin{document}


\title{Simple Lattice-Models of Ion Conduction:\\
       Counter Ion Model vs.\ Random Energy Model}

\author{J. Reinisch}
\author{A. Heuer}%

\affiliation{
Westf\"{a}lische Wilhelms-Universit\"{a}t M\"{u}nster\\
Institut f\"{u}r Physikalische Chemie and SFB 458\\
Schlossplatz 4/7, 48149 M\"{u}nster, Germany
}

\date{\today}

\begin{abstract}
The role of Coulomb interaction between the mobile particles in ionic conductors is
still under debate.
To clarify this aspect we perform Monte Carlo
simulations on two simple lattice models (Counter Ion Model and Random Energy Model)
which contain Coulomb
interaction between the positively charged mobile particles, moving on a
static disordered energy landscape. We find that
the nature of static disorder plays an important role if one
wishes to explore the impact of Coulomb interaction on the microscopic dynamics.
This Coulomb type
interaction impedes the dynamics in the
Random Energy Model, but enhances dynamics in the Counter Ion Model
in the relevant parameter range.
\end{abstract}

\maketitle

\section{\label{introduction}Introduction}
\begin{figure}[b]
  \includegraphics[width=8.6cm]{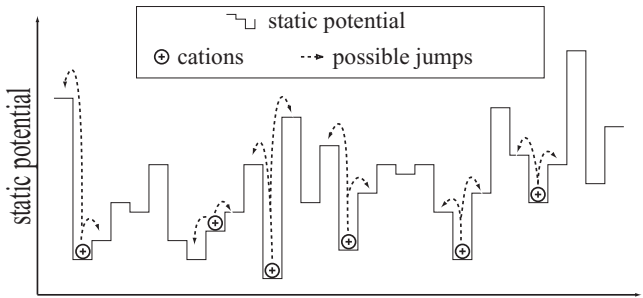}
  \caption{\label{figrescheme}One dimensional illustration of the static potential
                              landscape in the random energy model. The arrows
                              indicate possible jumps.}
\end{figure}

Many disordered insulating materials show universal
behavior in their ionic DC- and AC-conductivity.
Two prominent examples are the Arrhenius temperature dependence
of the DC-conductivity and the fact that AC-conductivity data for
different temperatures can be scaled onto a single master curve.
This curve is similar for most materials.
This observed universality \cite{Dyre:2000,Funke:2000}
has stimulated researches to find a common mechanism for ion conductivity in these materials.
A different type of system are disordered Anderson insulators, displaying electron transport.
They are denoted Coulomb glasses. In contrast to ion conductors non-localized dynamical
processes play an important role in these systems. The non-localized dynamics give rise to a
very different temperature dependence of the DC-conductivity at low temperatures, which turns
out to be proportional
to $exp[-(T_M/T)^{1/4}]$ (Mott`s law \cite{Mott:1968}) or to $exp[-(T_0/T)^{1/2}]$
 (Efros-Shklovskii law \cite{Efros:1975}).

Similar computational approaches have been chosen to investigate both types of problems
and various theoretical models have been developed in this context
\cite{Funke:1997,Summerfield:1984,Dyre:1993,Dyre:1993a,Pollak:1985,Shklovskii:1984}.
The focus of this work lies on the effect of Coulomb interaction on the microscopic dynamics
in ion conductors.
We chose two similar lattice models to investigate the relevant microscopic dynamics:
The Counter Ion Model (\cim)
\cite{Dietrich:1994, Dietrich:1996} and the Random Energy Model with cation-cation
interaction (\rem) \cite{Maass:1991}, which has some features in common with the Coulomb glass
model used for the second group of disordered solids
\cite{Pollak:1985,Shklovskii:1984}. The \cim\ and the \rem\ are designed to
reflect important
aspects of vitreous ion conductors, which are a high degree of disorder and mobile charged
ions
in a fixed glass network. The models differ in the way how the
time independent (static) disorder is realized. In this paper we analyze the effects of
Coulomb interaction among mobile ions and show that the nature of the disorder has great
influence on the dynamics.
In Sect. \ref{Models} we present the models as well as computational details, while in
the Sect. \ref{Results} the results are presented and discussed.

\section{Models and Computational details\label{Models}}
\subsection{\label{similar}Similarities of both models}
Both models base on a single type of mobile particles restricted to discrete sites in a
simple cubic lattice
with \var{l^3} sites of distance \var{a}.
The distance for a single jump is
limited to \var{a}, and occupied positions are forbidden. All mobile particles
have the same positive charge. In contrast to other works on the \cim\
\cite{Dietrich:1994, Dietrich:1996} the strength of the Coulomb interaction between the mobile
particles can be varied independently from the strength of static disorder
by appropriate selection of parameters, as it is also possible for the \rem;
see below for precise definitions.
This variation is essential to elucidate the effect of Coulomb interaction among the mobile
particles on their dynamics. There are
two energy contributions to the total Hamiltonian: Mobile particles moving on a
time dependent potential surface (generated by the particles themselves) and on a time
independent potential surface. We call the first dynamic and the latter static.
The dynamic part has the same form in both models and the Hamiltonian for the
cation-cation interaction is,
\begin{equation}
   H_{cat}=\frac{1}{2}\sum_{i\neq i'}\frac{n_i n_{i'} e^2}{r_{ii'}} .
\end{equation}
A configuration with sites $i$ is described by the occupation numbers $n_i=1$ for
occupied sites and
$n_i=0$ for empty sites. The omitted factor $4\pi\epsilon_0$ is taken in account by use of appropriate units.
The mean nearest-neighbor interaction in a system with randomly
distributed cations can be written as
\begin{equation}
    \vc=\frac{e^2}{r_s} .
\end{equation}
\rs\ is the mean distance of nearest neighbors in such a system with cation-concentration
$c=N/l^3$,
\begin{equation}
  \rs=\sqrt[3]{\frac{3}{4\pi c}} .
\end{equation}
Furthermore we introduce the dimensionless parameter \Gcat\ via
\begin{equation}
  \Gcat=\frac{\vc}{\kbt}=\frac{e^2}{\kbt r_s}.
  \label{eqngammadef}
\end{equation}
In the literature this parameter is already established for the \rem\ \cite{Maass:1991} and
we use it for the \cim\ to ensure comparability.
The Hamiltonian for the cation interaction can thus be rewritten as,
\begin{equation}
  \frac{H_{cat}}{\kbt}=\Gcat\cdot \frac{H_{cat}}{\vc} .
  \label{eqngamma}
\end{equation}
In what follows \vc\ is regarded as a constant for a given concentration.
Here $\Gcat\propto e^{2}$ is a measure for the interaction strength between the cations relative to \kbt.

\begin{figure}[t]
  \includegraphics[width=8.6cm]{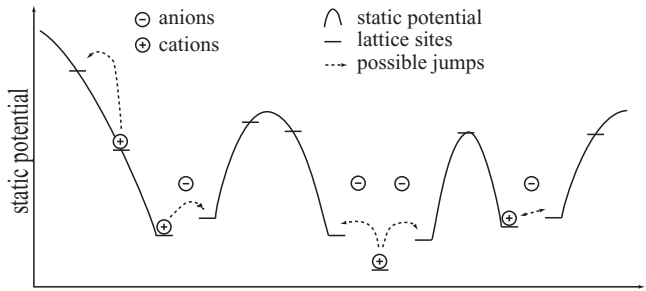}
  \caption{\label{figcischeme}One dimensional illustration of the static potential
                              landscape in the counter ion model. The arrows
                              indicate the possible jumps.}
\end{figure}
\begin{figure}
  \includegraphics[clip=,width=8.6cm]{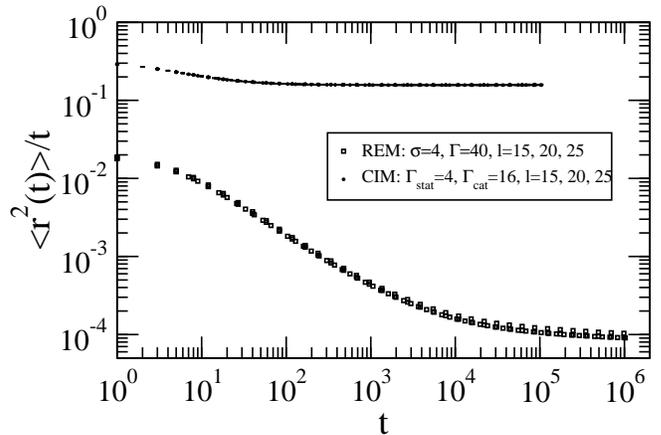}
  \caption{Typical errors and finite size effects for both models. The largest
           errors are of the size of the symbols and occur at large t. }
  \label{figerror}
\end{figure}

\subsection{Static Disorder in the Random Energy Model}
The \rem\ features a straightforward approach of the
principle of representing complexity by randomness. Each site \var{i}
gets a random energy taken from a Gaussian distribution with standard
deviation \sep\ and mean value $0$. The Hamiltonian for the static disorder becomes,
\begin{equation}
  H_{stat}=\sum_{i}n_i\cdot E_{i}^{static} .
\end{equation}
The model is fully determined by two dimensionless parameters: \s\ is connected to the
standard deviation of the random site energies by,
\begin{equation}
  \s=\frac{\sep}{\kbt} .
\end{equation}
In changing \s\ one can change the strength of static disorder relative to temperature
in the system.
As before, the second parameter \Gcat\ determines the cation-cation interaction strength.
The total Hamiltonian reads,
\begin{equation}
  \frac{H}{\kbt}=\Gcat\cdot\frac{H_{cat}}{\vc}+\frac{H_{stat}}{\kbt} .
\end{equation}
In \figref{figrescheme} the potential landscape of the \rem\ is illustrated.

\subsection{Static Disorder in the Counter Ion Model}
In the \cim\ static disorder is generated by randomly placing negatively charged
particles at centers of cubic lattice cells. As for cations these
anions cannot occupy the same site. The resulting potential landscape
(see \figref{figcischeme}) is different from that seen in \figref{figrescheme}.
As for the \rem\ we used \Gcat\ to control the interaction strength among cations
and defined a parameter \Gan\ adequate to control the interaction strength between cations
and anions. It is defined in analogy to \Gcat\ as,
\begin{equation}
  \Gan=\frac{k_{ca}\cdot\vc}{\kbt}.
\end{equation}
The factor \var{k_{ca}} has been introduced to vary the strength of the cation-anion
interaction separately from the cation-cation interaction. For $k_{ca}=1$ and thus
$\Gcat=\Gan$ one recovers the standard choice of parameters for the \cim.
In contrast to previous works we wish
to break with charge neutrality and rather interpret $H_{stat}$ as a general static disordered
energy landscape. Thus \Gan\ can be varied in analogy to \s\, thereby modifying the strength
of the disorder.
The total Hamiltonian for the \cim\ is,
\begin{equation}
  \frac{H}{\kbt}=\Gcat\cdot\frac{H_{cat}}{V_{c}}+\Gan\cdot \frac{H_{stat}}{\vc} .
\end{equation}
\subsection{Computational Settings}
The number of lattice sites in one dimension was set to 20 for all data throughout
this work. For simulating bulk properties we apply periodic boundary conditions
and the minimum image convention \cite{Metropolis:1953}. As shown in
in \figref{figerror} the smallness of the finite size effects justifies our choice of
20 lattice sites per dimension.
All presented data were calculated with a cation concentration of $0.03$.
The number of anions for generating the static energy landscape in the \cim\ was
kept identical to the number of cations. The contributions of the Coulomb interaction
were calculated via Ewald summation.
The number of different starting configurations ranges from 5 to 10.
This rather
small number made it possible to simulate a large range of parameters.
For model systems with $c=0.03$ and $e^2=1$, which were used throughout this paper,
one has $\vc=0.501$.

\section{Results\label{Results}}
\begin{figure}
  \includegraphics[clip=,width=8.6cm]{REddcofsandslopefig4}
  \caption{Left: Dependency of the low frequency diffusion constant \ddc\ on \s\ in the \rem.
           The lines
           correspond to exponential fits, the data for $\s=0$ are
           omitted. The dotted line describes the temperature dependency of \ddc\ for a
           constant quotient $\frac{\s}{\Gcat}=\frac{\sep}{\vc}=0.05$ . The errors are
           smaller than the symbol size.\\
           Right: Dependency of the slope on \Gcat. The data match the theoretical limit at $\Gcat=0$.}
  \label{figddcofs}
\end{figure}
\begin{figure}
  \includegraphics[clip=,width=8.6cm]{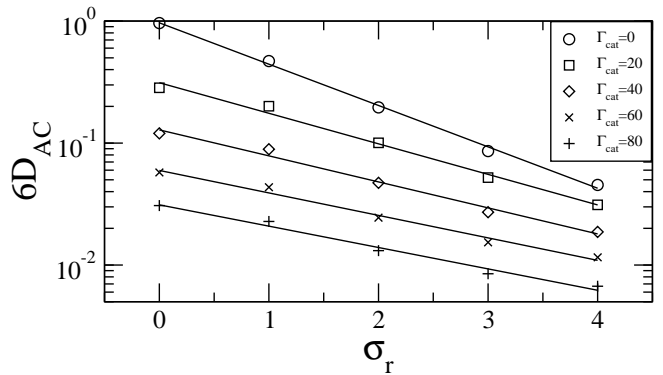}
  \caption{Dependency of the high frequency diffusion constant \dac\ on \s\ in the \rem.
           The lines are exponential fits and serve as guides to the eye.
           The errors are within the symbol size.}
  \label{figdacofs}
\end{figure}
The diffusion constants are taken from \rsdt-data. \figref{figerror}\
shows some examples. Each curve begins with a short-time
plateau changes into a dispersive regime from \var{t_1} to \var{t_2} and forms a
long-time plateau beyond \var{t_2}. In the dispersive regime between \var{t_1}
and \var{t_2} the curve follows a Jonscher type \cite{Jonscher:1983} power law
\begin{equation}
  \langle r^2(t)\rangle/(6t) \sim t^{k'-1}
\end{equation}
The following relations are used to determine the diffusion constants,
\begin{eqnarray*}
 &6D(t)= \rsdt ,\\
 &6\ddc=\lim_{t\rightarrow\infty},\rsdt ,\\
 &6\dac=\lim_{t\rightarrow 0}\rsdt.
\end{eqnarray*}
The high frequency diffusion constant \dac\ was also calculated from the average
hopping rate $a^2\langle w \rangle=6\dac$. For details on the general behavior of the
\rem\ and the \cim, see \cite{Maass:1991,Dietrich:1996, Dietrich:1998}.

In the \rem\ the presence of cation-cation interaction leads to a significant decrease in diffusion
and an increase in dispersion \cite{Maass:1991}. A quantitative analysis of the behavior
can be found in \cite{Maass:1996}. For a \rem\ without long range Coulomb
interaction, i.e.\ $\Gcat=0$, the activation energy \ea\ can be calculated as the difference between the
Fermi energy and the critical percolation energy \cite{Dyre:1993, Maass:1996}. For a
concentration of $0.03$ the Fermi energy is $-1.88\sep$ and the percolation energy
is $-0.49\sep$. Hence the activation energy is $-0.49\sep+1.88\sep=1.39\sep$. This reasoning
is only valid for low temperatures $\s\gg 1$. For high temperatures $\s\ll 1$ it can be shown
\cite{Maass:1996} that the activation energy is roughly equal to
$\sep/\sqrt{\pi}=0.56\sep$.
Here we analyze the \rem\ including long range Coulomb interaction for which this analytical
treatment is no
longer possible. Simulated data for various \Gcat\ and \s\ are shown
in \figref{figddcofs}(left).
The data for $\Gcat=0$ correspond to vanishing cation-cation interaction and the regression
for the low
temperature regime leads to an activation energy of $1.41\sep$ which is in good agreement with
the theoretical value resulting from percolation arguments, see above. One can
easily see that the low temperature part of all curves can be approximated fairly well by
straight lines. The first data point for $\s=0$ is omitted for all \Gcat. Interestingly, the
change of slope at high temperatures, i.e.\ small \s, as predicted for $\Gcat=0$, seems to
hold for large cation-cation interaction, too. The data shown in
\figref{figddcofs}(left) can be fitted with the activation energy dependent on \s, \Gcat\ and a
cross term $\Gcat\cdot\s$,
\begin{equation}
  \ln[6\cdot\ddc]
  =-(a_{1}\s+a_{2}\Gcat+a_{3}\s\Gcat)\label{eqnansatz}
\end{equation}

The slope of the fits in \figref{figddcofs}(left) can be expressed according to \eqnref{eqnansatz}
as $a_{1}+a_{3}\Gcat$. \figref{figddcofs}(right) illustrates this
dependency. The three coefficients turn out to be $a_{1}=1.41$, $a_{2}=0.063$ and $a_{3}=0.015$.
The non-vanishing value of \var{a_{3}} is of particular interest since it
contradicts the conclusions of Maass et al. \cite{Maass:1996}. They showed that for c=0.01
the low temperature behavior is activated, i.e.\ $\ln[6D]\propto\frac{1}{T}$ when varying the
temperature. Since $\s\propto\frac{1}{T}$ and $\Gcat\propto\frac{1}{T}$ this statement is
identical to $a_{3}=0$. We have included a curve in \figref{figddcofs}(left) which represents a
variation of the temperature only, this curve shows a proportionality to $A/T+B/T^2 (A,B>0)$.
In contrast to this result experimental data
for low temperatures
show no curvature \cite{Maass:1996} or opposite curvature \cite{Namikawa:1974}.
This temperature dependence is also very different to that observed in Coulomb glass for which,
if at all, a description with $B<0$ would be appropriate for a limited temperature regime
in order to recover the Efros-Shklovskii law.

\begin{figure}
  \includegraphics[clip=,width=8.6cm]{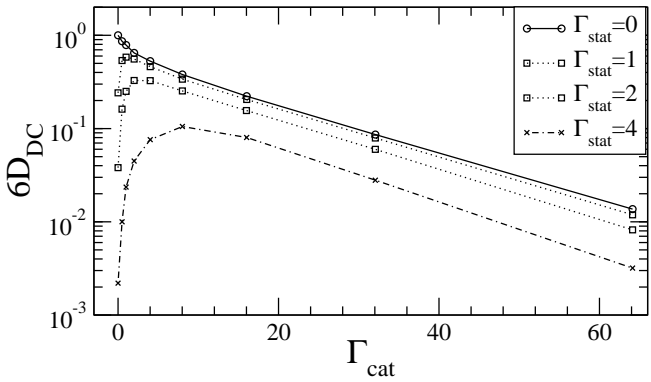}
  \caption{\label{figciddcofg} Dependency (\cim) of the low frequency diffusion constant \ddc\
           on \Gcat. The maximum (\Gcatmax) shifts to larger \Gcat\ for higher values of \Gan.
           The errors are smaller than the symbols.}
\end{figure}
\begin{figure}
  \includegraphics[clip=,width=8.6cm]{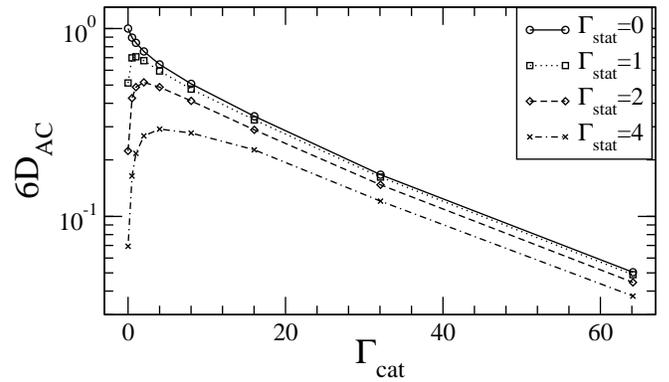}
  \caption{\label{figcidacofg} Dependency (\cim) of the high frequency diffusion constant \dac\
           on \Gcat. The maximum (\Gcatmax) shifts to larger \Gcat\ for higher values of \Gan.
           The errors are smaller than the symbols.}
\end{figure}
\begin{figure}
  \includegraphics[clip=,width=8.6cm]{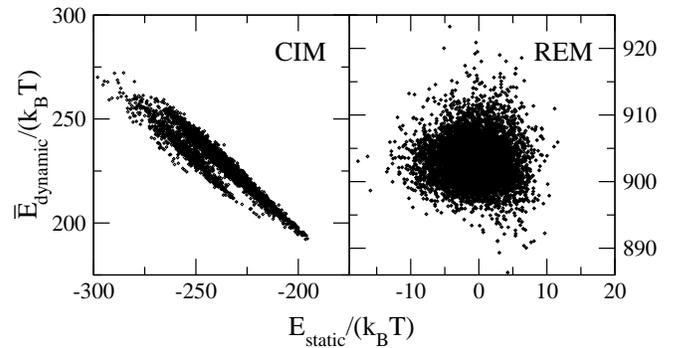}
  \caption{\label{figkorr}Correlation between static potential and
           mean value of the dynamic potential for the \cim\ (left; $\Gcat=10$; $\Gan=10$)
           and the
           \rem\ (right; $\Gcat=40, \s=4$). Each site is represented by a
           single diamond symbol ($20^3$ for each graph).
           In the \cim\ a strong correlation is observed, the slope is close to one.
           In the \rem\ no significant correlation is observed.}
\end{figure}
\begin{figure}
  \includegraphics[clip=,width=8.6cm]{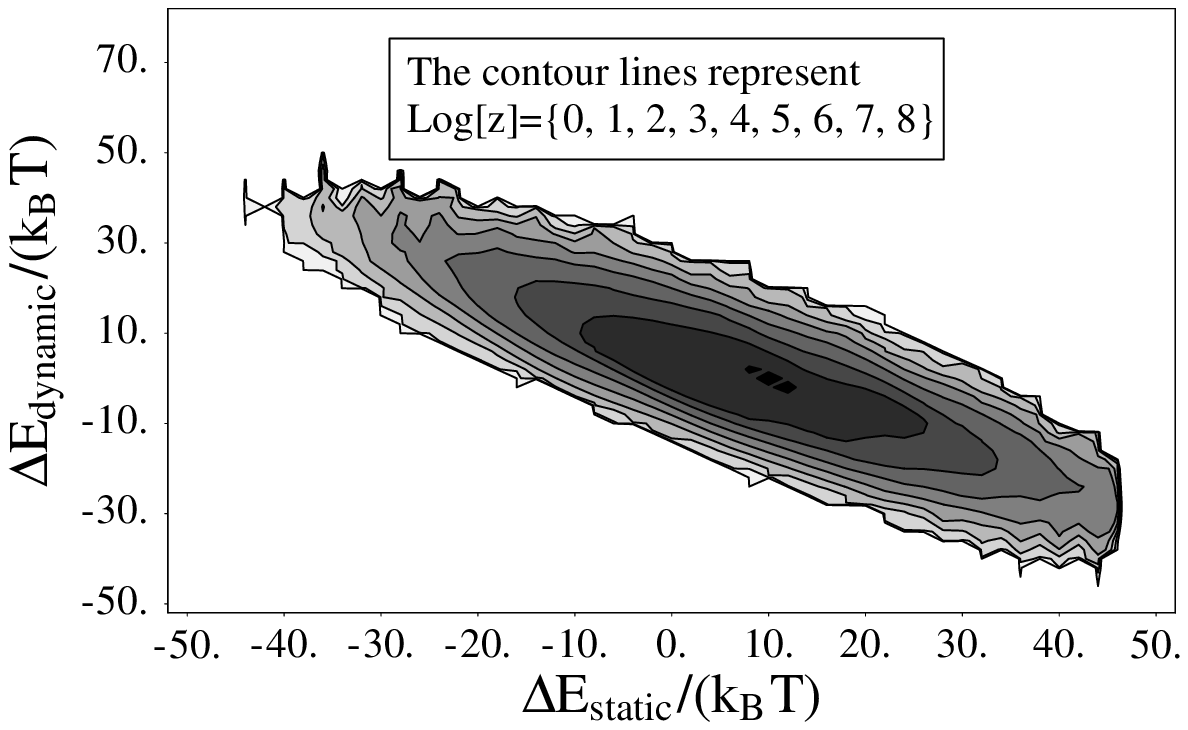}

  \includegraphics[clip=,width=8.6cm]{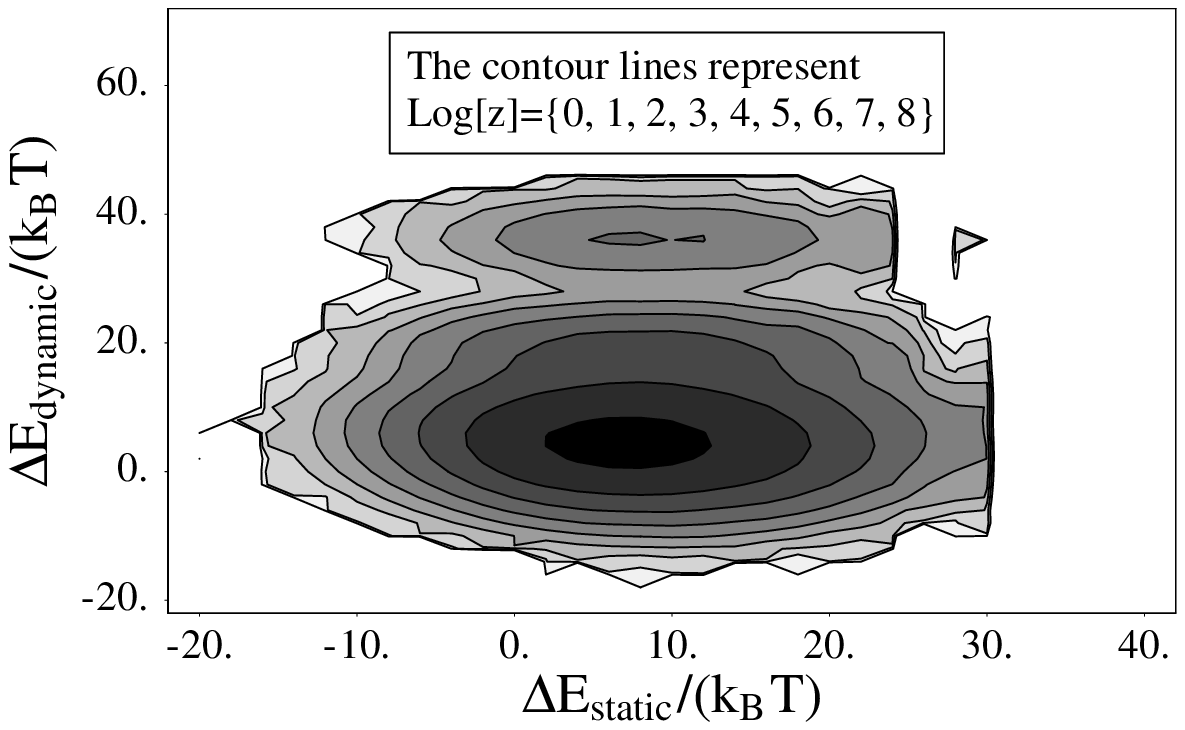}
  \caption{\label{figsprung}Matrices of successful and unsuccessful trial jumps
           in the
           \cim\ (top, $\Gcat=\Gan=10$) and the \rem\ (bottom, \s=4, \Gcat=40)
           resolved for static and dynamic
           jump energies. \var{z} is a counter of how many jumps happen per
           energy interval during a simulation run. The choice of parameters has
           no influence on the qualitative picture.}
\end{figure}

Furthermore we also analyzed the dependency of \dac\ on \Gan\ and \Gcat.
\figref{figdacofs} shows data for \dac\ in analogy to those in
\figref{figddcofs}(left). Exponential fits are not as accurate as for \ddc. Though the data
cannot be accurately fitted as straight lines an approximated slope is decreasing
with increasing \Gcat. Thus \dac\ has a different behavior as compared to \ddc.

The analysis for the \cim\ was performed in a similar way. The first surprising
feature is the increase in mobility of the cations for \ddc\ (Fig.\ \ref{figciddcofg}) and
\dac\ (Fig.\ \ref{figcidacofg}) when increasing \Gcat\ from $0$ with fixed \Gan. This
is the completely opposite behavior as compared to the
\rem. Increasing \Gcat\ further leads to an increase in diffusion until a maximum is reached,
from this maximum the diffusion decreases and the slope becomes constant in a logarithmic plot.
Obviously, a simple functional form as \eqnref{eqnansatz} can not be found for the \cim.
A difference between \ddc\ and \dac\ is that \Gcatmax\ for \ddc\ is approximately equal to \Gan\
whereas \Gcatmax\ for \dac\ increases more rapidly with \Gan.

The different behavior of the \rem\ as compared to the \cim\ can be rationalized by a
closer inspection of the differences in the potential surfaces.
Comparing
\figref{figrescheme} and \figref{figcischeme} and taking the definitions of the models
into account it's evident that the static potential of a lattice site in the \cim-surface is
spatially correlated, whereas in the \rem\ no correlation among adjacent sites exists.
This correlation in the
\cim\ has a direct consequence:
\figref{figkorr} illustrates a
correlation of the static potential of a lattice site with the mean value of the
dynamic potential. For a single starting configuration the static energy of each site
is constant during simulation, being denoted as $E_{static}/\kbt$ in \figref{figkorr}.
The mean value
of the dynamic energy $\bar{E}_{dynamic}$ at some site is generated by averaging over all
energies which a particle at this site has due to $H_{cat}$
during a simulation.
The observed correlation in \figref{figkorr}(left) indicates
a correlation of $H_{cat}$ and $H_{an}$; high cation-cation energy corresponds to
low static energy and vice versa.
It could be shown that the three regions observed in the \cim\ correspond to three
different types of lattice sites. There are sites with no, one
or two adjacent anions, i.e.\ anions with a distance of $\sqrt{3}\cdot \frac{a}{2}$.
No such correlation is observed for the \rem.

To elucidate the
impact of this correlation on the ion dynamics all occurring trial jumps,
accepted or not by the Metropolis algorithm, were
recorded
during a MC simulation with respect to their static (\des) and dynamic (\ded) jump energy differences
(\figref{figsprung}). The sum of \des\ and \ded\
is the total energy difference for a jump. One apparent effect of the correlation for the
\cim\ is quite
obvious: a high energy contribution from the static energy is accompanied by a
low energy contribution from the dynamic energy and
vice versa. Therefore to first approximation the total energy does not change during a jump
giving rise to faster dynamics.
The \rem\ does not show this behavior. For the \rem\ both contributions are independent
and both impede the cation dynamics. A consequence is a different jump energy
distribution in the \cim\ as compared to the \rem.
\begin{figure}
  \includegraphics[clip=,width=8.6cm]{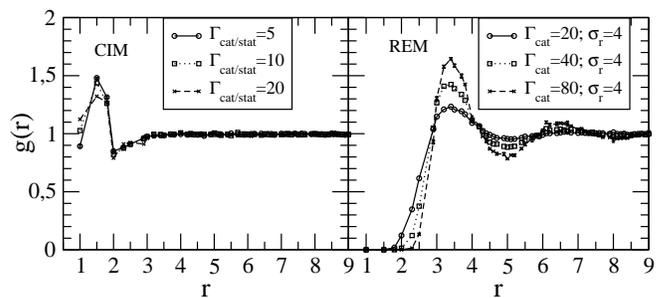}
  \caption{Radial density function \gofr\ for the \cim(left) and the \rem(right). For the
           \cim\ the parameters are chosen to show the temperature dependency and for the
           \rem\ the dependency of \gofr\ on \Gcat.
           The errors are within the size of the symbols.}
  \label{figgofrboth}
\end{figure}
To further explore the reasons for this behavior the
radial density function \var{g(r)} of both models is illustrated in \figref{figgofrboth}.
Obviously, the cations in the \rem\ have a structure, they prefer a certain distance to each
other which returns periodically in the graph. This structure is very similar to that found
in simple liquid systems. The \cim\ does not show such a structure, instead it shows
an unusual
high concentration of cations in very short distances and beyond $r\approx 3$ the bulk
density is
already reached. The found structures give possible explanations for the
observed
behaviors: In the \rem\ each cation is surrounded by a cage of other cations, and this cage
slows down the
movement \cite{Heuer:1998} because most moves would increase \var{H_{cat}}.
In the \cim\ the cations prefer to
populate low energy positions around
the anions, which results in an effective screening of the anion charge. This screening flattens
the overall energy landscape and thus gives rise to increased mobility. Increasing
\Gcat\ reduces the clustering due to cation-cation repulsion but also supports screening
due to higher cation charge. These two effects compete and may account for the observed behavior,
see \figref{figciddcofg} and \figref{figcidacofg}.

We are now able to reason the parameter dependency of \eqnref{eqnansatz} for the \rem.
The first term arises from the static disorder of the energy landscape. The second term is
introduced by a cage effect as mentioned above. The cross term on the other hand has no single
microscopic origin for \ddc\ and \dac. For \ddc\ it
may arise from relaxation of the system to a small perturbation (i.e. a cation jump).
By a jump process a cation may have left a well-adjusted environment
and is surrounded by an energetically unfavorable environment. Apart from jumping back
the total system may also adjust to this new situation as already
formulated in the concept of mismatch and
relaxation by Funke \cite{Funke:1997, Funke:2000}. The situation at the new position improves
energetically after some time due to the subsequent relaxation of the adjacent particles,
thereby reducing the probability of the backjump with time. In the presence of disorder this
neighbor relaxation is, of course, much slower since also the neighbors experience the
effect of the static disorder. Therefore it is more likely for the central particle to jump
back, giving rise to a decrease of \ddc\ due to the simultaneous effect of static disorder
and the cation-cation interaction.
For \dac\ an increase in \Gcat\ reduces the dependency on \s\ and therefore the above argumentation
fails. For no cation interaction and in the limit of zero temperature the cations occupy
only the $N$ lowest
sites in static energy. By introducing cation interaction the emerging liquid structure
forces the particles to occupy also some sites which are higher in static energy
(but lower in total energy), this would lead to the observed reduction in \s\ dependence.

\section{Summary\label{Summary}}

The \rem\ and the \cim\ display major qualitative differences in their
dependences
of \dac\ and \ddc\
on the system parameters. We gave qualitative reasoning by taking a
closer view into the microscopic
origins as effects of caging, relaxation, screening and disorder were
observed.
We have also shown that the conduction in the \rem\ does not show simple
Arrhenius behavior and despite their similarities the \rem\ behaves
quite differently as
compared to the Coulomb glass due to the different nature of trial moves.
The results of this work give rise to a new important variable in
disordered solids: The spatial correlation of the potential energy
hypersurface. Currently we
are analyzing the energy hypersurface of real ion conductors to extract
this important piece of information.

\begin{acknowledgments}
We wish to acknowledge the support of the SFB 458 as well as helpful discussions with
R. Banhatti, P. Maass and B. Roling. We acknowledge additional support by B. Roling
for providing us the Ewald summation program.
\end{acknowledgments}
\bibliography{RHpaper}

\end{document}